\documentclass[a4paper,11pt]{article}
\pdfoutput=1
\usepackage{jcappub}
\usepackage{amssymb}
\usepackage{amsmath}
\usepackage{mathrsfs}
\usepackage{bm}
\usepackage{graphicx}
\usepackage{hyperref}
\usepackage{booktabs}

\newcommand{\cm}{\mathrm{cm}}
\newcommand{\pc}{\mathrm{pc}}
\newcommand{\kpc}{\mathrm{kpc}}

\newcommand{\keV}{\mathrm{keV}}

\newcommand{\GeV}{\mathrm{GeV}}
\newcommand{\TeV}{\mathrm{TeV}}
\newcommand{\romN}[1]{\uppercase\expandafter{\romannumeral #1\relax}}

\begin{document}

\title{Constraints on a mixed model of dark matter particles and primordial black holes from the Galactic 511 keV line}

\author[a,b,c]{Rong-Gen Cai,}
\author[a,b]{Yu-Chen Ding,}
\author[a,b,1]{Xing-Yu Yang,\note{Corresponding author.}}
\author[a,b,c]{Yu-Feng Zhou}

\affiliation[a]{
    CAS Key Laboratory of Theoretical Physics, Institute of Theoretical Physics, Chinese Academy of Sciences, P.O. Box 2735, Beijing 100190, China
}
\affiliation[b]{
    School of Physical Sciences, University of Chinese Academy of Sciences, No.19A Yuquan Road, Beijing 100049, China
}
\affiliation[c]{
    School of Fundamental Physics and Mathematical Sciences, Hangzhou Institute for Advanced Study, University of Chinese Academy of Sciences, Hangzhou 310024, China
}

\emailAdd{cairg@itp.ac.cn}
\emailAdd{dingyuchen@itp.ac.cn}
\emailAdd{yangxingyu@itp.ac.cn}
\emailAdd{yfzhou@itp.ac.cn}

\abstract{
    The galactic 511 keV gamma-ray line has been observed since 1970's, and was identified as the result of electron-positron annihilation, but the origin of such positrons is still not clear.
    Apart from the astrophysical explanations, the possibilities that such positrons come from dark matter (DM) annihilation are also widely studied.
    Primordial black hole (PBH) is also an extensively studied candidate of DM.
    If PBHs exist, the DM particles may be gravitationally bound to the PBHs and form halo around PBHs with density spikes.
    DM annihilation in these density spikes can enhance the production rate of positrons from DM particles, but the signal morphology is similar to the decaying DM.
    We consider such a mixed model consisting of DM particles and PBHs and obtain the upper limit from the data of 511 keV gamma-ray line from INTEGRAL/SPI on the decaying component of DM particles and the constraint on the PBH abundance.
    These constraints are general and independent of particle DM models.
    For the mixed model consisting of excited DM and PBHs, the constraints on the PBH abundance can be down to $O(10^{-17})$ for DM particle with mass around $1~\TeV$, which is more stringent than that obtained from the extragalactic gamma-ray background.
}

\maketitle

\section{Introduction}

In the standard cosmology model, the ordinary baryonic matter only contributes about 5\% of the energy budget of the universe, while the rest of the universe, in which 26\% is dark matter (DM) and 69\% is dark energy (DE), is barely known~\cite{Aghanim:2018eyx}.
There are many DM candidates, including astrophysical objects such as primordial black holes (PBHs), and particles beyond the standard model of particle physics~\cite{Bertone:2016nfn}.

The PBHs are formed from the gravitational collapse of the overdense regions in the early universe~\cite{Zeldovich:1967,Hawking:1971ei,Carr:1974nx}, and have recently attracted considerable attention (see review papers \cite{Sasaki:2018dmp,Carr:2020gox}, and references therein).
There are plenty of scenarios that lead to PBH formation~\cite{Liu:2019lul,Cai:2019bmk}, and most of them require a mechanism to generate large overdensities.
These overdensities are often of inflationary origin, and will collapse if they are larger than a certain threshold when reentering the horizon~\cite{Stewart:1996ey,Stewart:1997wg,Leach:2000ea,Cheng:2018yyr,Gao:2018pvq,Germani:2017bcs,Nakama:2018utx}.
Except for being candidate of DM, PBHs can also act as the seeds for galaxy formation~\cite{Bean:2002kx,Kawasaki:2012kn,Nakama:2017xvq,Carr:2018rid}, and the sources of gravitational waves of LIGO/Virgo detection~\cite{Bird:2016dcv,Sasaki:2016jop}.

In principle either PBH or DM particle can account for the total DM abundance, but there is no requirement that one of them must dominate the total abundance.
Indeed, a mixed DM model consisting of DM particles and PBHs can lead to many interesting consequences compared with a single component DM model, even if PBHs only constitute a small fraction of the DM.
Such a mixed DM model has been studied before, and has been constrained by the gamma-ray background observed by Fermi-LAT~\cite{Eroshenko:2016yve,Boucenna:2017ghj,Adamek:2019gns}.
It can also be constrained by radio searches with the Square Kilometer Array (SKA) and gravitational wave searches with LIGO/Virgo and the future Einstein Telescope (ET)~\cite{Bertone:2019vsk}.

The 511 keV gamma-ray line was first detected at the galactic center and was identified as the result of electron-positron annihilation in 1970's~\cite{Johnson.172.1972.feb,Haymes.201.1975.nov,Leventhal.A225.1978.09}.
There are also many recent measurements such as the SPI spectrometer on the INTEGRAL observatory (INTEGRAL/SPI)~\cite{Siegert:2015knp} and the COSI balloon telescope~\cite{Kierans:2019aqz}.
This gamma-ray line is mostly due to parapositronium annihilation of thermal or near-thermal positrons~\cite{Churazov:2004as,Jean:2005af}, and the absence of continuous high-energy spectrum from positron annihilation in flight implies that the initial energy of these positrons is less than a few MeV~\cite{Beacom:2005qv}.
Most plausible sources of positrons are believed to be distributed in the disk of the Galaxy, but this gamma-ray line displays a puzzling morphology with bright emission from an extended bulge region and faint disk emission.
Early analyses suggested the flux ratio of bulge-to-disk to be around $1 \sim 3$~\cite{Knodlseder:2005yq}, and more recent analyses give a smaller value around $0.6$~\cite{Bouchet:2010dj,Siegert:2015knp}.
Positrons injected from the decay of isotopes coming from nucleosynthesis in stars can tentatively explain the emission from the galactic disk~\cite{Prantzos:2010wi,Bartels:2018eyb}, but the origin of the emission in the bulge is still under debate.

Many sources have been proposed to explain the origin of the bulge 511 keV gamma-ray line, such as the $\beta^{+}$ decay of stellar nucleosynthesis products (e.g. $^{26}$Al, $^{44}$Ti and $^{56}$Ni)~\cite{Milne:1999ky,Prantzos:2010wi,Crocker:2016zzt}, the low-mass X-ray binaries~\cite{Bartels:2018eyb} and the neutron star mergers~\cite{Fuller:2018ttb}.
Apart from those astrophysical explanation, given the existence of DM in our galaxy, scenarios that the bulge 511 keV gamma-ray line originated from DM have also been extensively investigated~\cite{Boehm:2003bt,Boehm:2004gt,Finkbeiner:2007kk,Pospelov:2007xh,Vincent:2012an,Cudell:2014jba,Farzan:2017hol}.
It is possible that the low-energy galactic positrons are produced by direct annihilation of light dark matter (LDM) (about few MeV) particles into electron-positron pairs~\cite{Boehm:2003bt}, or by the excited dark matter (XDM) model.
In the XDM model, the DM particles can be heavy, excited states of heavy DM are produced in collisions between DM particles in the ground state, and the electron-positron pairs are produced through the decay of the excited state into the ground state~\cite{Finkbeiner:2007kk,Pospelov:2007xh}.
It was shown that these DM models can be constrained by Cosmic Microwave Background (CMB) observations~\cite{Frey:2013wh} and the light WIMP (lighter than around 10 MeV) explanation of the 511 keV line was claimed to be excluded~\cite{Wilkinson:2016gsy}.
However, there are still models of DM which can avoid these constraints~\cite{Lawson:2016mpu,Farzan:2017hol}, and the recent refined analysis found that LDM with mass down to around 1 MeV can be made consistent with CMB and Big Bang Nucleosynthesis (BBN) observations~\cite{Sabti:2019mhn,Ema:2020fit}.

The DM scenarios that galactic low-energy positrons originated from the interaction of two DM particles such as direct annihilation or collision, will give a morphology of 511 keV gamma-ray line which depends on the squared density of DM particles $\rho^{2}(\bm{r})$, while the DM scenarios that positrons produced from processes like decay of DM particles will give a morphology which depends on the density $\rho(\bm{r})$.
Therefore the decaying DM scenarios will lead to a more spread distribution of 511 keV signal than annihilating scenarios, and it was found that the data favor the annihilating DM scenarios~\cite{Ascasibar:2005rw,Abidin:2010ea,Vincent:2012an}.
However this dose not rule out subdominant decaying components, since the analyses of 511 keV data in earlier studies are performed only for single-component DM scenarios.

With the existence of PBHs, the DM particles may be gravitationally bound to the PBHs and form halos around PBHs with density spikes~\cite{Eroshenko:2016yve}.
Since the interaction of DM particles is related to the particle density, the formation of density spikes can leave some imprints in the 511 keV gamma-ray line observations.
Therefore, by analysing the data of 511 keV gamma-ray line, one can get constraints on both PBHs and DM particles.\footnote{We focus on relatively large mass PBHs, PBHs with small mass can also be constrained by 511 keV line as positrons can be produced by its Hawking radiation~\cite{DeRocco:2019fjq,Laha:2019ssq,Dasgupta:2019cae,Laha:2020ivk}.}
As we will show, as to the contribution on the morphology of 511 keV signal, the halo of one PBH is equivalent to one decaying particle, thus the PBH abundance can be constrained stringently by the upper limit of the decaying components.

A DM model with both annihilating and decaying components should be considered.  In this article we consider a mixed model consisting of PBHs and self-annihilating DM particles.
The PBHs attract DM particles to create density spikes, and contribute to galactic gamma-ray production as a decaying like component.
We use 511 keV gamma-ray data from INTEGRAL/SPI to constrain the contribution of such a component, and give an upper limit on the abundance of PBHs.
These constraints are general and independent of particle DM models, which means that any particle DM model proposed to explain the 511 keV gamma-ray line observations will always give such constraints on the abundance of PBHs.
For the mixed model consisting of XDM and PBHs, the constraints on PBH abundance for DM particle with mass around $1~\TeV$ can be down to $O(10^{-17})$, which is much more stringent than that obtained from the extragalactic gamma-ray background.

This paper is organized as follows.
In scetion~\ref{sec:model}, we introduce the DM model consisting of both particle DM and PBHs and explain how it can explain and be constrained by the 511 keV gamma-ray observation.
Then we prepare the ingredient for such constrains by performing a bayesian inference on 511 keV gamma-ray data from INTEGRAL/SPI in section~\ref{sec:data} and derive the DM halo profile around PBHs in section~\ref{sec:halo}.
Finally in section~\ref{sec:constraint} we give the model independent constraints on the PBH abundance for various masses of DM particle and PBH.  The final section is devoted to conclusions.

\section{511 keV gamma-ray line from the mixed DM model}\label{sec:model}

There are many DM scenarios proposed to explain the 511 keV gamma-ray line, which can be classified into annihilating and decaying scenario.
In these scenarios low energy positrons are produced from the DM annihilation/decay in the DM halo, and annihilate with electrons into 511 keV gamma-rays.
For a self-annihilating DM model, the positron production rate from the DM annihilation is given by
\begin{equation} \label{eq:nea}
    \dot{n}_{e^{+}}= n_{\chi}^{2} \langle \sigma v \rangle / 2,
\end{equation}
where $n_{e^{+}}$ and $n_{\chi}$ are the number densities of positrons and DM particles, respectively, $\langle \sigma v \rangle$ is the thermally averaged cross-section for DM annihilating into positrons.\footnote{The factor should be $1/4$ instead of $1/2$ if the DM is not self-annihilating.}
In the decaying scenario, the positron production rate from the DM decay is given by
\begin{equation}\label{eq:neD}
    \dot{n}_{e^{+}}=n_{\chi} \Gamma,
\end{equation}
where $\Gamma$ is the decay rate.

Most of the annihilating elctron-positron pairs form positronium ~\cite{Harris:1998tt,Kinzer:2001ba,Churazov:2004as}, with a positronium formation fraction $f_{\mathrm{p}}\approx 0.967$~\cite{Jean:2005af}.
In this process, $1/4$ of the annihilation take place in the parapositronium state yielding 2 photons with $E=511~\keV$ which accounts for the observations, while the remaining $3/4$ yield 3 photons with $E< 511~\keV$.
For these positrons which do not form positronium, they annihilate directly into 2 photons with $E=511~\keV$.  Thus the total number density of 511 keV photons produced per unit time is related to the positron production rate,
\begin{equation}
    \dot{n}_{\gamma}=2((1-f_{\mathrm{p}})+f_{\mathrm{p}}/4) \dot{n}_{e^{+}}.
\end{equation}

Given a specific DM halo model for the Milky Way, one can compute the intensity distribution of 511 keV signature as a function of galactic longitude $l$ and latitude $b$, by integrating the emissivity $\dot{n}_{\gamma} (\bm{r})$ along the line of sight (l.o.s.), which is
\begin{equation}
    I(l,b)=\frac{1}{4\pi}\int_{\mathrm{l.o.s.}} \dot{n}_{\gamma}(\bm{r}) ds,
\end{equation}
with the distance to the Galactic center
\begin{equation}
    r=\sqrt{r_{\odot}^{2}+s^{2} - 2 r_{\odot} s \cos b \cos l},
\end{equation}
where $s$ denotes the distance from the solar system along the integrated line of sight, and $r_{\odot} \approx 8.5~\kpc$ is the distance from the Sun to the Galactic center~\cite{Camarillo:2017vlt}.
Thus the total flux of the 511 keV photons from DM is given by the integral of the intensity distribution,
\begin{equation}
    \Phi=\int I(l,b) d\Omega.
\end{equation}

Suppose the density profile of Milky Way DM halo is $\rho(\bm{r})$.
For the annihilating DM scenarios, the intensity distribution of its 511 keV signature is
\begin{equation}
    I_{\mathrm{A}}(l,b)
    =2(1-\frac{3}{4}f_{\mathrm{p}}) \frac{1}{4\pi}
    \int_{\mathrm{l.o.s.}} \frac{1}{2}
    \frac{\langle \sigma v \rangle}{m_{\chi}^{2}} \rho^{2}(\bm{r}) ds
    \text{,}
\end{equation}
while for the decaying scenarios, it is
\begin{equation}
    \label{eq:I_decay}
    I_{\mathrm{D}}(l,b)
    = 2(1-\frac{3}{4}f_{\mathrm{p}})\frac{1}{4\pi}
    \int_{\mathrm{l.o.s.}} \frac{\Gamma}{m_{\chi}} \rho(\bm{r}) ds
    \text{.}
\end{equation}
Note that the intensity distribution is $I_{\mathrm{D}} \sim \rho(\bm{r})$ for decaying scenarios, while $I_{\mathrm{A}} \sim \rho^{2}(\bm{r})$ for annihilating scenarios, thus the decaying scenarios will lead to a more spread distribution compared with the annihilating scenarios.

Except for particle DM, PBH is also an extensively studied candidate of DM.
There are many experimental constraints on the fraction of PBHs in the DM, it is possible that  PBHs with mass in the range $10^{-16} \sim 10^{-12}~M_{\odot}$ constitute all the DM, but for the steller mass PBHs, the constraints suggest that PBHs are subdominant to the rest of DM~\cite{Carr:2020gox}.
The fraction of PBHs in DM is defined as
\begin{equation}
    f_{\mathrm{PBH}}\equiv \rho_{\mathrm{PBH}}/\rho_{\mathrm{DM}},
\end{equation}
so the corresponding fraction of DM particles is $\rho_{\chi}=(1-f_{\mathrm{PBH}})\rho_{\mathrm{DM}}$.
In this work, we consider a mixed DM model where the DM particle component is dominated, namely, $f_{\mathrm{PBH}} \ll 1$.

Consider the PBH formation in the radiation-dominated era, once the DM particles have kinetically decoupled from the primordial plasma, they could be gravitationally bound to the PBHs and form halo with density spikes.
Since we only need that PBHs are formed prior to the kinetic decoupling of DM particles from the primordial plasma, we are not assuming any specific formation mechanism of PBHs.

Denoting the number of positrons produced from the halo of \emph{one} PBH per unit time by $\Gamma_{\mathrm{PBH}}$, then the positron production rate from \emph{all} the halo around PBHs is given by
\begin{equation}\label{eq:nePBH}
    \dot{n}_{e^{+}}=n_{\mathrm{PBH}}\Gamma_{\mathrm{PBH}},
\end{equation}
where $n_{\mathrm{PBH}}$ is the number density of PBHs.
Comparing eq.~\eqref{eq:nePBH} with eq.\eqref{eq:neD}, we can find that \emph{the halo of one PBH} is equivalent to \emph{one decaying particle}, this is the key point for constraining PBH abundance with 511 keV gamma-ray line data.

We consider the mixed model consisting of annihilating DM particles and PBHs, which will give a conservative constraint since if there are additional decaying particle components in DM, the PBH abundance will be constrained more stringently.
In such a mixed DM model, the intensity distribution of 511 keV signature from unbounded DM particles and halo of PBHs are
\begin{equation}
    I_{\mathrm{A}}(l,b) = 2(1-\frac{3}{4}f_{\mathrm{p}})\frac{1}{4\pi}
    \int_{\mathrm{l.o.s.}} \frac{1}{2} \frac{\langle \sigma v
    \rangle}{m_{\chi}^{2}} (1-f_{\mathrm{PBH}})^{2}\ \rho^{2}(\bm{r}) ds
\end{equation}
and
\begin{equation}\label{eq:D_PBH}
    I_{\text{`D'}}(l,b) = 2(1-\frac{3}{4}f_{\mathrm{p}})\frac{1}{4\pi}
    \int_{\mathrm{l.o.s.}} \frac{\Gamma_{\mathrm{PBH}}}{M_{\mathrm{PBH}}}
    f_{\mathrm{PBH}}\ \rho(\bm{r}) ds
\end{equation}
respectively, where $m_{\chi}$ and $M_{\mathrm{PBH}}$ are the mass of DM particles and PBHs, and
\begin{equation}
    \Gamma_{\mathrm{PBH}}=\int dr^{3} \frac{1}{2} \frac{\langle \sigma v
    \rangle}{m_{\chi}^{2}} \rho^{2}_{\chi, \mathrm{PBH}} ,
\end{equation}
where $\rho_{\chi, \mathrm{PBH}}$ is the density profile of DM halo around PBHs.
Noticing that there are only annihilating particles and PBHs in the DM, however, as we have explained that the halo of one PBH is equivalent to one decaying particle, which can also be verified by comparing eq.~\eqref{eq:D_PBH} with eq.~\eqref{eq:I_decay}, therefore the notation $I_{\text{`D'}}$ means that it is equivalent to decaying DM when one only cares the morphology, but it is not the decaying DM.

Define
\begin{align}
    \label{eq:CA}
    \frac{\langle \sigma v \rangle}{m_{\chi}^{2}} (1-f_{\mathrm{PBH}})^{2} =
    C_{\mathrm{A}} ,\\
    \label{eq:CD}
    \frac{\Gamma_{\mathrm{PBH}}}{M_{\mathrm{PBH}}} f_{\mathrm{PBH}} =
    C_{\mathrm{D}} ,
\end{align}
then we have
\begin{equation}\label{eq:con_fPBH}
    \frac{f_{\mathrm{PBH}}}{(1-f_{\mathrm{PBH}})^{2}} = \frac{2
    M_{\mathrm{PBH}}}{\int dr^{3} \rho_{\chi,\mathrm{PBH}}^{2}}
    \frac{C_{\mathrm{D}}}{C_{\mathrm{A}}} .
\end{equation}
The ratio $C_\mathrm{D}/C_\mathrm{A}$ can be inferred from the morphology of 511 keV gamma-ray observations.
Along with the density profile of DM halo around PBHs , eq.~\eqref{eq:con_fPBH} gives constraints on the abundance of PBHs.

\section{Fit to INTEGRAL/SPI data}\label{sec:data}

It is found that the morphology of the observed 511 keV gamma-ray line distribution is concentrated in the galactic center, favoring an annihilating DM scenario.
While the predicted flux distribution in a decaying DM scenario is too flat to be used to explain the observation~\cite{Ascasibar:2005rw,Abidin:2010ea,Vincent:2012an}.
However, this does not necessarily exclude the possibility where decaying DM components exist along with dominant annihilating DM components, and their existence can be constrained by the morphology of 511 keV gamma-ray line observation.

In addition to the DM produced positrons, we also consider the contribution of $\beta^+$ emission from radioactive isotopes including ${}^{26}$Al and ${}^{44}$Ti in the galactic disk.
The distribution of such emissions can be described by Robin young stellar disk (YD) model~\cite{Robin:2004qd,Knodlseder:2005yq} as
\begin{equation}
    I_{\mathrm{YD}}(l,b) =
    2(1-\frac{3}{4}f_{\mathrm{p}})\frac{1}{4\pi}
    \int_{\mathrm{l.o.s.}} \dot{n}_{\mathrm{YD}} \left[
        e^{- \left( \frac{a} {R_{0}} \right)^{2}} - e^{- \left( \frac{a}{R_{i}}
    \right)^{2}} \right] ds ,
\end{equation}
with
\begin{equation}
    a^{2}=x^{2}+y^{2}+R_0^2{z^2}/{z_0^2} \mathrm{,}
\end{equation}
where $x,y,z$ are coordinates with coordinate origin in galaxy center and $x$-$y$ plane in galaxy disk, $R_0$ is the disk scale radius, $R_i$ is the truncation radius and $z_{0}$ is the vertical scale height.
Therefore the intensity distribution of 511 keV signature in this analysis can be written as
\begin{equation} \label{eq:flux-model}
    I(l,b) =  I_{\mathrm{A}}(l,b) + I_{\text{`D'}}(l,b) + I_{\mathrm{YD}}(l,b).
\end{equation}

We perform a bayesian inference over INTEGRAL/SPI data to determine the parameters appeared in eq.~\eqref{eq:flux-model}.
The three free parameters ($C_{\mathrm{A}}$, $C_{\mathrm{D}}$ and $\dot{n}_\mathrm{YD}$) are set with uniform priors in linear space.
The shape of the disk emission profile is fixed as $R_0=5~\kpc$, $R_i=3~\kpc$ and $z_0=125~\pc$, which matches the distribution of 1809 keV gamma-ray line from ${}^{26}$Al decay~\cite{Diehl:2005py}.
For $\rho(\bm{r})$, two DM halo profiles, Einasto~\cite{Einasto:2009zd} and NFW~\cite{Navarro:1996gj}, are considered separately as
\begin{align}
    \rho(r)_{\text{Einasto}}
  &= \rho_\odot \exp \left( \frac{2}{\alpha} \left[
          \left( \frac{r_\odot}{r_{s}} \right)^{\alpha}
          - \left( \frac{r}{r_{s}} \right)^{\alpha}
  \right] \right)
  , \\
  \rho(r)_{\text{NFW}}
  &= \rho_\odot \frac {r_\odot}{r}
  \left( \frac {r_\odot + r_s} {r + r_s} \right)^2 ,
\end{align}
where $\alpha=0.17$, $r_{s}=20~\kpc$ and the local DM density $\rho_\odot = 0.3~\GeV/\cm^3$.
The INTEGRAL/SPI data of diffusive Galactic emission used in this analysis is extracted from \cite{Bouchet:2011fn}, where the gamma-ray intensities are averaged over each latitude interval of each bin and over longitude $|l| < 23.1~\mathrm{deg}$.
The flux is integrated over the energy range $E = 200-600~\keV$.
The uncertainties of the data are treated as gaussian errors.
Thus the likelihood $L$ can be related to the $\chi^2$ as
\begin{equation} \label{eq:chi2}
    -2 \ln L = \chi^2 =
    \sum_i^N \left( \frac{ I_i - \hat{I}_i } { \sigma_i } \right)^2 ,
\end{equation}
where $I_i$ and $\sigma_i$ are the central values and uncertainties of gamma-ray intensity measured for each bin, and $\hat{I}_i$ is the corresponding averaged intensity predicted by models described as eq.~\eqref{eq:flux-model}, e.g.,
\begin{equation}
    \hat{I}_i =
    \frac { \int_{\mathrm{bin}_i} I(l,b) d\Omega }
    { \int_{\mathrm{bin}_i} d\Omega }.
\end{equation}

\begin{figure}[htpb]
    \centering
    \includegraphics[width=0.49\textwidth]{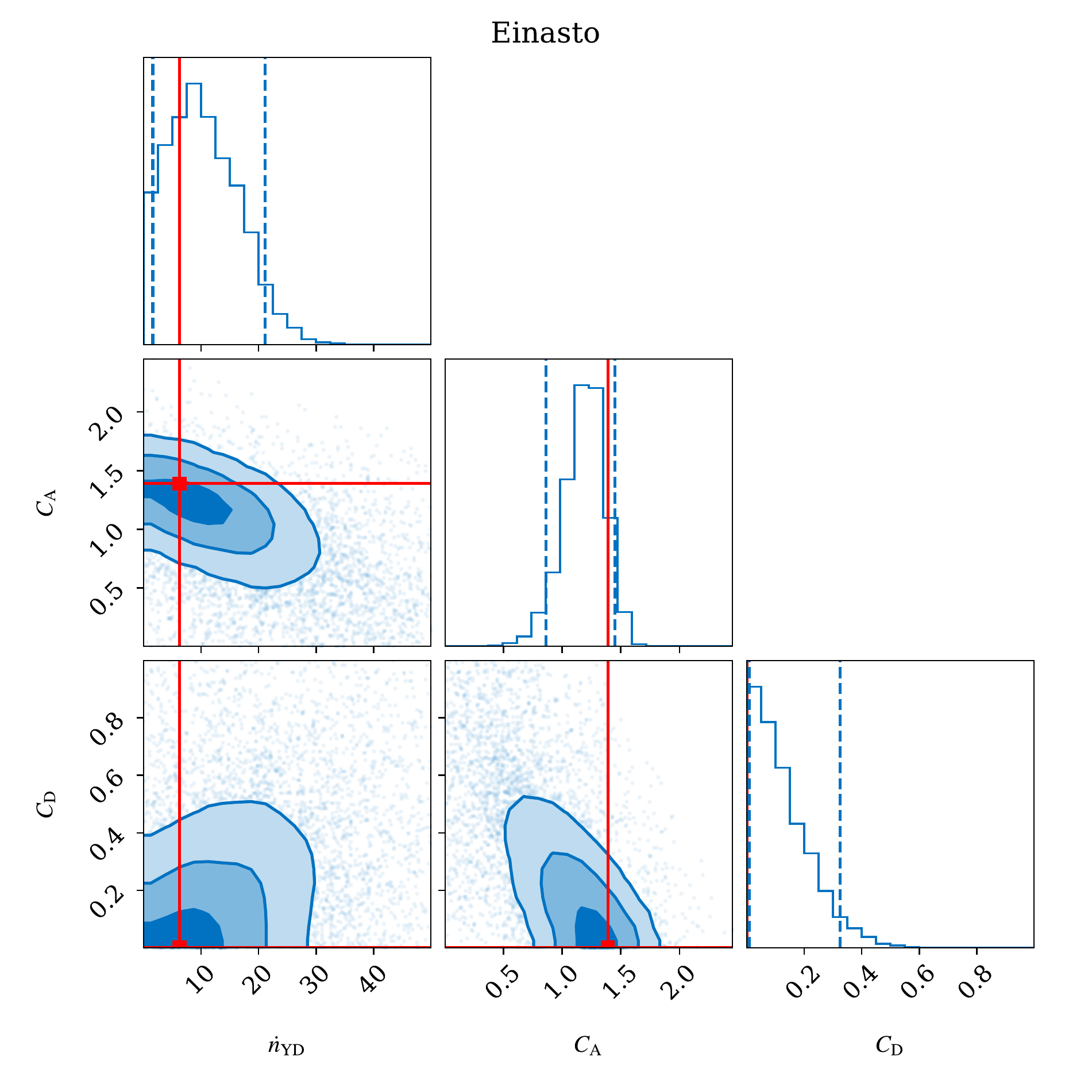}
    \includegraphics[width=0.49\textwidth]{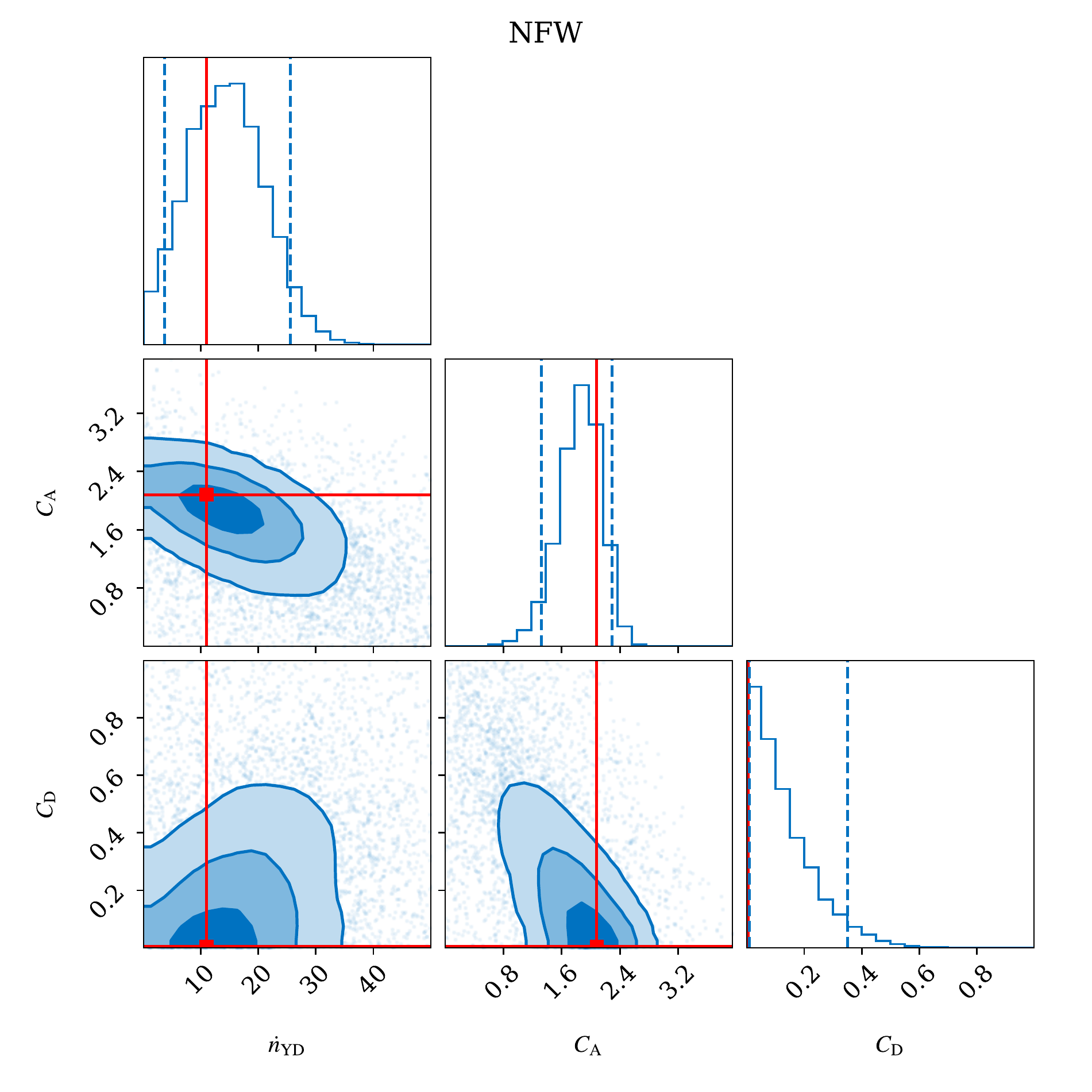}
    \caption[]{Triangle plots of posteriors given by fitting 511 keV gamma-ray data from INTEGRAL/SPI measurement.
        The results for Einasto (left panel) and NFW (right panel) profiles are shown.
        Three different contour levels in the plots are 0.39, 0.86 and 0.99, which corresponds to 1-, 2- and 3-$\sigma$ level in 2D distributions.
        The red solid lines indicate the best-fit parameter values that give minimum $\chi^2$ defined as eq.~\eqref{eq:chi2}.
        The blue dashed lines indicate the 5\% and 95\% quantiles for each parameter.
        The units of the quantities can be found in the caption of table~\ref{tab:result}.
        These figures are generated by {\sc corner.py}~\cite{corner}. }
    \label{fig:triangles}
\end{figure}

\begin{figure}[htpb]
    \centering
    \includegraphics[width=0.7\textwidth]{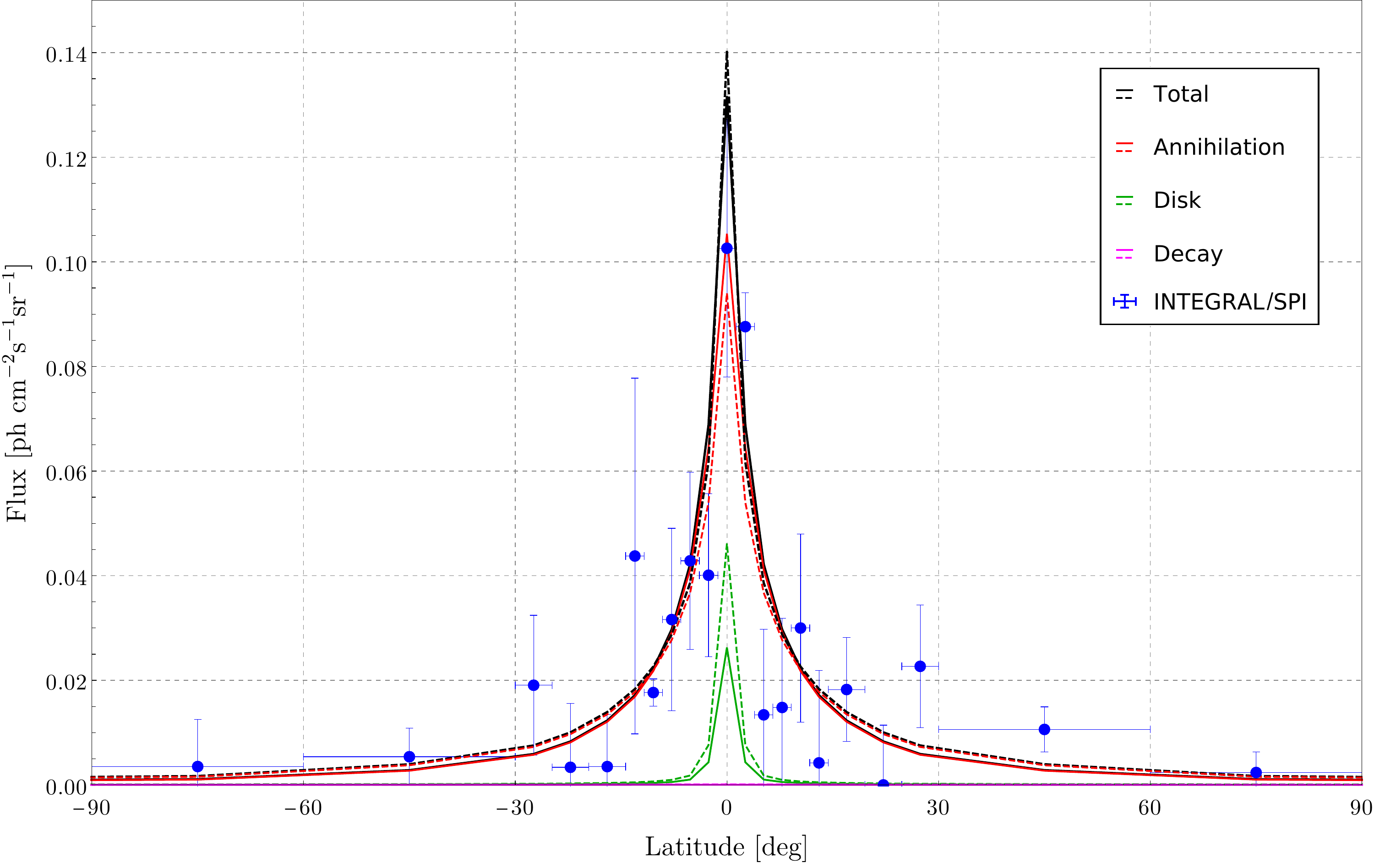}
    \caption[]{Predicted 511 keV fluxes with different profiles compared with INTEGRAL/SPI data.
        Solid (dashed) lines correspond to the best-fit parameters for Einasto (NFW) profile presented in table~\ref{tab:result}.
        The decay component of DM does not contribute to the total gamma-ray flux since $C_{\mathrm{D}}$ vanishes in both results. }
    \label{fig:fluxes}
\end{figure}

\begin{figure}[htpb]
    \centering
    \includegraphics[width=0.7\textwidth]{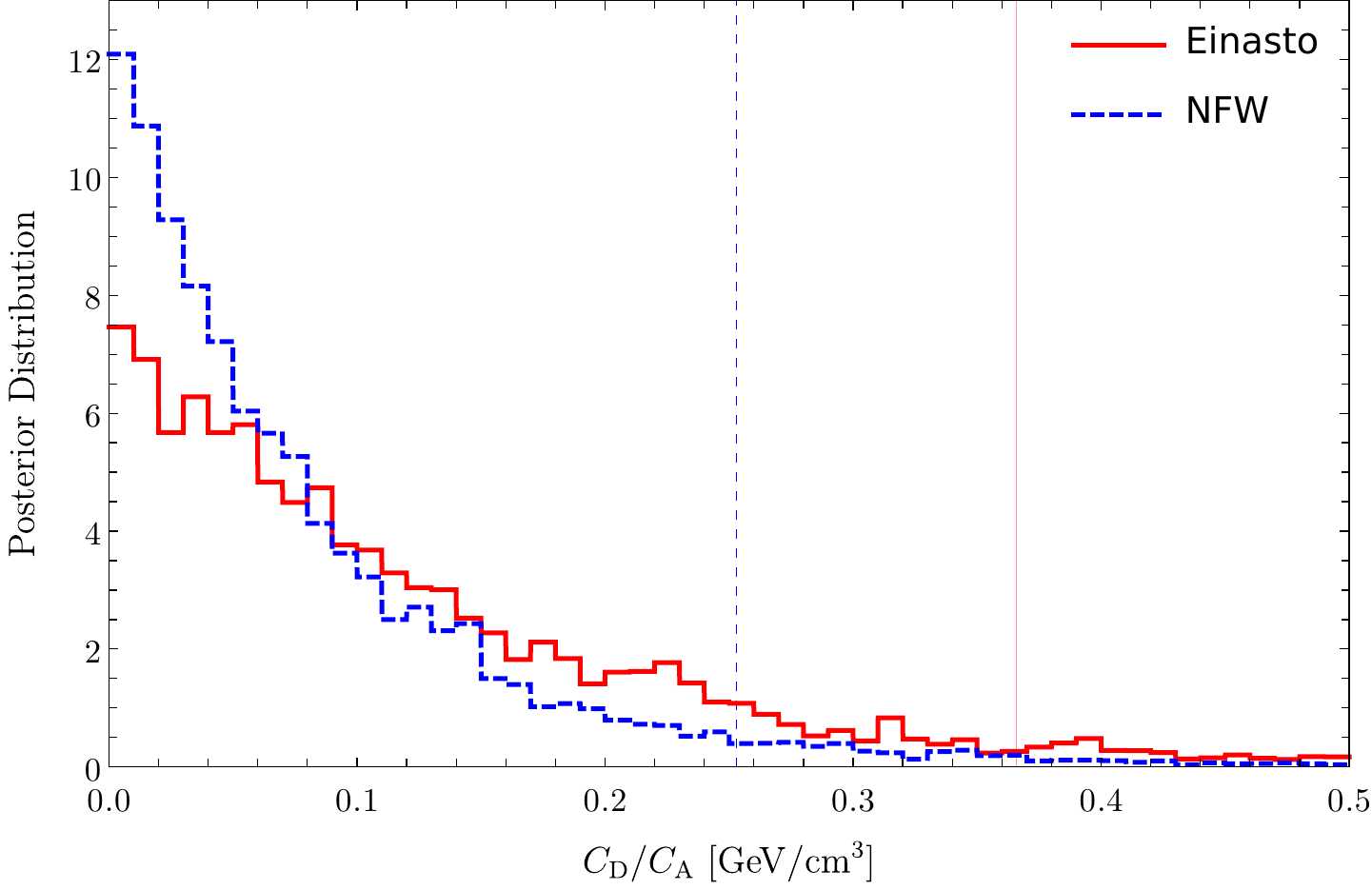}
    \caption[]{Posterior distributions of $C_\mathrm{D}/C_\mathrm{A}$.
        Verticle lines mark the 95\% upper limits of $C_\mathrm{D}/C_\mathrm{A}$ with Einasto (red solid) and NFW (blue dashed) profiles. }
    \label{fig:cdca}
\end{figure}

With the likelihood defined in eq.~\eqref{eq:chi2}, the posterior distributions of the three-dimentional parameter space can be numerically sampled using MultiNest~\cite{Feroz:2013hea,Feroz:2008xx,Feroz:2007kg}.
The marginalized 2D joint distributions and 1D posteriors are shown in figure~\ref{fig:triangles}.
It can be seen that for both DM profiles $\dot{n}_\mathrm{YD}$ and $C_\mathrm{A}$ have gauss-like posteriors, while the most probable value for $C_\mathrm{D}$ converges to zero, with a 95\% upper limit of $~ 3.3\times 10^{-24} \ \GeV^{-1} \mathrm{s}^{-1}$.
The favored values for the parameters and their confidence intervals are briefly listed in table~\ref{tab:result}.
With the best-fit model parameters, the predicted 511 keV fluxes are plotted in figure~\ref{fig:fluxes} along with data.
We also show the ratio of $C_\mathrm{D}$ and $C_\mathrm{A}$ in table~\ref{tab:result} and its posterior distribution as figure~\ref{fig:cdca}, which will be used to constrain the abundance of PBH in eq.~\eqref{eq:con_fPBH}.
\begin{table}[htpb]
    \centering
    \caption[]{Summary of the fitting results for 511 keV data of INTEGRAL/SPI.
        For each DM profile, the first row shows the best-fit value of each parameter, corresponding to the $\chi^2$ value shown in the second column.
        The second row of each DM profile shows the means and standard variances for parameters with non-zero best-fit value, and 95\% upper limits for parameters whose best-fit value converges to zero.
        Units for $\dot{n}_\mathrm{YD}$, $C_\mathrm{A}$ and $C_\mathrm{D}$ are $10^{-23} \cm^{-3} \mathrm{s}^{-1}$, $10^{-23} \GeV^{-2} \cm^3 \mathrm{s}^{-1}$ and $10^{-23} \GeV^{-1} \mathrm{s}^{-1}$, respectively. }
    \label{tab:result}
    \begin{tabular}{lccccc}
        \toprule \midrule
        Profile &
        $\chi^{2}/\mathrm{d.o.f.}$ &
        $\dot{n}_{\mathrm{YD}}$ &
        $C_\mathrm{A}$ & $C_\mathrm{D}$ & $C_\mathrm{D}/C_\mathrm{A}$ \\
        \midrule
        Einasto & $29.9/18$ & 6.26 & 1.39 & 0 & 0 \\
          & & $10.5\pm6.13$ & $1.18\pm0.18$ & $<0.33$ & $<0.37$ \\
          \midrule
        NFW & $35.5/18$ & 11.0 & 2.08 & 0 & 0 \\
          & & $14.4\pm6.61$ & $1.84\pm0.30$ & $<0.35$ & $<0.25$ \\
          \midrule \bottomrule
    \end{tabular}
\end{table}

Comparing the two results for different DM profiles, the one with Einasto profile has a better goodness of fit than that with NFW profile, while their best-fit values and confidence regions are consistent without a significant divergence.
In both cases, the value for $C_{\mathrm{D}}$ converges to zero, which is consistent with the statement from earlier works \cite{Ascasibar:2005rw,Vincent:2012an} that a decaying DM component is disfavored by the observed morphology of 511 keV gamma-ray.

We notice that the fitting result of $C_\mathrm{A}$ is larger by about an order of magnitude than those given in \cite{Ascasibar:2005rw,Vincent:2012an} which were obtained by fitting galactic 511 keV line with only annihilating DM.
This discrepancy can be explained by the data used in our analysis.
The integrated energy range of the data we used ($200-600~\keV$) are larger than that used in those works.
This leads to a higher flux by about an order of magnitude, resulting in a larger estimated values of both $C_\mathrm{A}$ and $C_\mathrm{D}$.
However, it is $C_\mathrm{D}/C_\mathrm{A}$ that matters when one constrains the PBH abundance, where the discrepancy is canceled.

\section{Density profile of halo around PBHs}\label{sec:halo}

For a DM particle at position $\tilde{r}$ with velocity $\tilde{\bm{v}}$, it would spend a fraction $2dt/\tau_{\mathrm{orb}}$ of its period at distances between $r$ and $r+dr$,\footnote{Due to the symmetry of the orbit, the particle passes the same radius twice, which leads to the factor of 2~\cite{Eroshenko:2016yve}.} where $\tau_{\mathrm{orb}}$ is the period of the particle's orbital motion around the PBH and $dt$ is the time it takes for the particle to move from $r$ to $r+dr$.
Suppose the PBH forms at time $\tilde{t}$, given the initial density of DM particles $\rho_{\mathrm{ini}}$, at later time $t>\tilde{t}$, the density of DM particle halo around PBHs $\rho_{\mathrm{b}}(r)$ can be written as the relation~\cite{Eroshenko:2016yve,Boucenna:2017ghj}
\begin{equation}
    \rho_{\mathrm{b}}(r) 4\pi r^{2} dr=\int 4\pi \tilde{r}^{2}
    d\tilde{r}\rho_{\mathrm{ini}}(\tilde{r}) \int d^{3}\tilde{\bm{v}}
    f_{\mathrm{B}}(\tilde{\bm{v}}) \frac{2 dt}{\tau_{\mathrm{orb}}} ,
\end{equation}
which follows from the Liouville equation and expresses the density conservation law in phase space integrated over the momenta by taking into account the volume transformation in momentum space, where the velocity distribution of DM particles $f_{\mathrm{B}}(\tilde{\bm{v}})$ is chosen to be Maxwell-Boltzmann distribution,
\begin{equation}
    f_{\mathrm{B}}(\tilde{\bm{v}}) = \frac{1}{(2\pi \bar{\sigma}^{2})^{3/2}} \exp \left( -\frac{\tilde{\bm{v}}^{2}}{2\bar{\sigma}^{2}} \right)
\end{equation}
with $\bar{\sigma}\equiv\sqrt{T/m_{\chi}}$ and the temperature of the primordial plasma $T$.
Therefore, we have
\begin{equation}\label{eq:rho_b}
    \rho_{\mathrm{b}}(r)=\frac{1}{r^{2}}\int d\tilde{r}\ \tilde{r}^{2}\
    \rho_{\mathrm{ini}}(\tilde{r}) \int
    d^{3}\tilde{\bm{v}}\ f_{\mathrm{B}}(\tilde{\bm{v}})
    \frac{2}{\tau_{\mathrm{orb}}} \frac{dt}{dr} .
\end{equation}

\begin{figure}[htpb]
    \centering
    \includegraphics[width=0.7\textwidth]{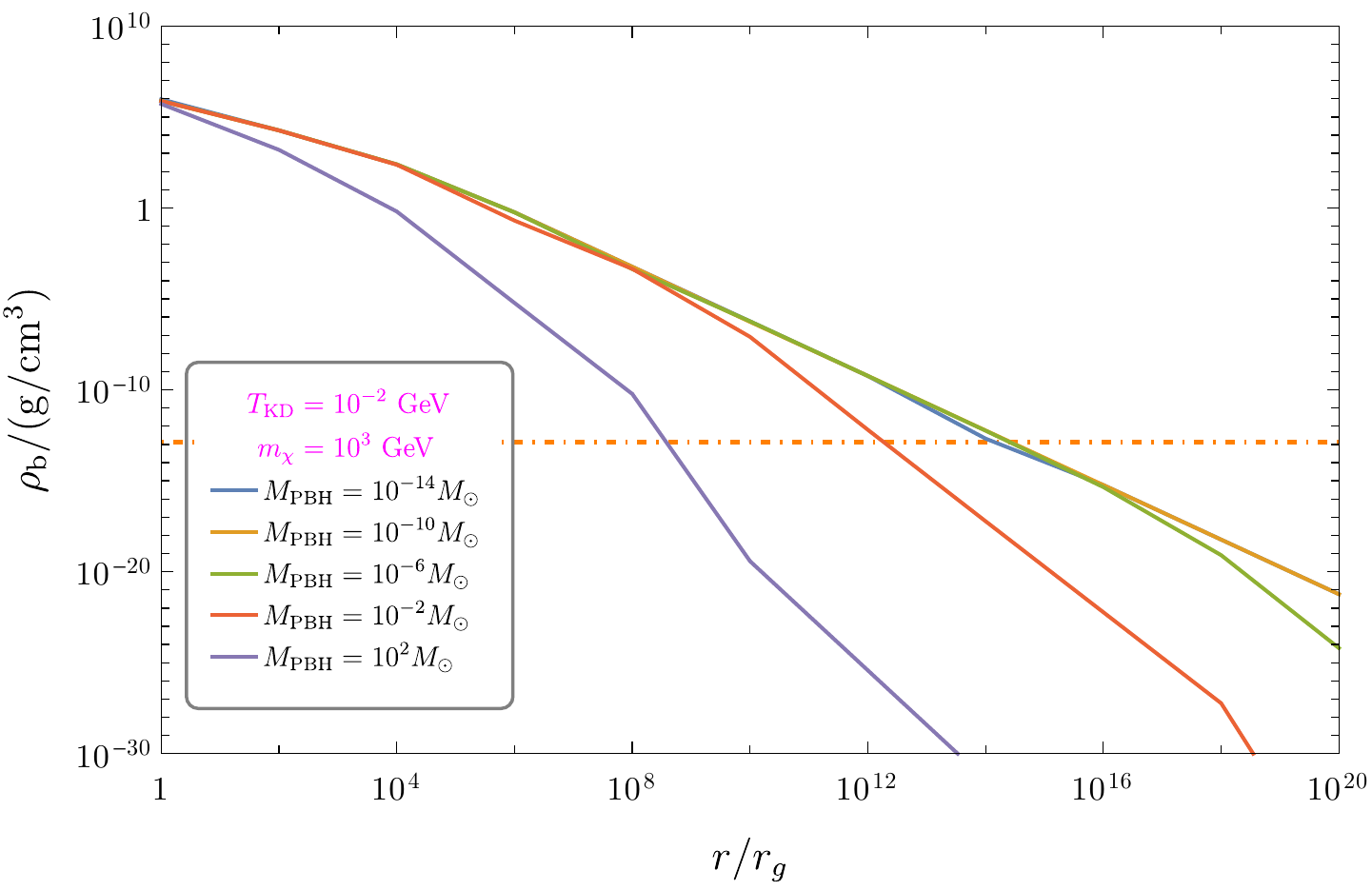}
    \caption[]{The density profile $\rho_{\mathrm{b}}$ of DM particles bound to
        a PBH with different values of $M_{\mathrm{PBH}}$ for $T_{\mathrm{KD}}=10^{-2}~\GeV$ and $m_{\chi}=10^{3}~\GeV$.
        For a given $T_{\mathrm{KD}}$ and $m_{\chi}$, the density profile for the lighter PBH constitutes an envelope to the profile for the heavier PBH, these profiles overlap in the small $r/r_{g}$ range and diverge in the large  $r/r_{g}$ range.
        The horizontal dot-dashed line denotes the maximum possible density at present time of the annihilating DM particles computed from eq.~\eqref{eq:rho_max}. }
    \label{fig:rho_b}
\end{figure}

In figure \ref{fig:rho_b}, we show the density profile $\rho_{\mathrm{b}}$ of DM particles bound to a PBH as a function of the rescaled radius $r/r_{g}$ (where $r_{g} \equiv 2G M_{\mathrm{PBH}}$) with DM particles whose mass $m_{\chi}=10^{3}~\GeV$ and temperature of kinetic decoupling $T_{\mathrm{KD}}= 10^{-2}~\GeV$.
The density profile for the lighter PBH constitutes an envelope to the profile for the heavier PBH, this is because the maximum rescaled radius that a PBH can gravitationally affect, i.e. $(3 M_{\mathrm{PBH}}/4\pi\rho_{r})^{1/3}/r_{g}$ (where $\rho_{r}$ is the energy density of the radiation-dominated universe), is smaller for the heavier PBH.

For the annihilating DM particles, their density will decrease with time, and there will be a maximum possible density at present time which is given by~\cite{Bringmann:2011ut}
\begin{equation}\label{eq:rho_max}
    \rho_{\mathrm{max}}=\frac{m_{\chi}}{\langle \sigma v \rangle_{\mathrm{a}} t_{0}},
\end{equation}
where $t_{0} \approx 4.3 \times 10^{17} \mathrm{s}$ is the age of the universe~\cite{Aghanim:2018eyx}, and $\langle \sigma v \rangle_{\mathrm{a}}$ is the thermally averaged cross-section for annihilation of DM particles, which is chosen to be $\langle \sigma v \rangle_{\mathrm{a}} \sim 3 \times 10^{-26}\ \mathrm{cm^{3} s^{-1}}$ in order to match the observed relic density.\footnote{Note that $\langle \sigma v \rangle_{\mathrm{a}}$ is different from $\langle \sigma v \rangle$ which only accounts for the channel of DM to positron.}
Therefore the density profile of DM particle halo around PBH is
\begin{equation}\label{eq:rho_chi_PBH}
    \rho_{\chi,\mathrm{PBH}}(r)=\min[\rho_{\mathrm{max}}, \rho_{\mathrm{b}}(r)],
\end{equation}
with critical radius $r_{c}$ satisfying $\rho_{\mathrm{b}}(r_{c}/r_{g})=\rho_{\mathrm{max}}$.
As shown in figure \ref{fig:rho_b}, $r_{c}/r_{g}$ increases as $M_{\mathrm{PBH}}$ decreases and when $M_{\mathrm{PBH}}$ smaller than a certain value, $r_{c}/r_{g}$ will be about constant.

\section{Constraints on PBHs}\label{sec:constraint}

\begin{figure}[htpb]
    \centering
    \includegraphics[width=0.49\textwidth]{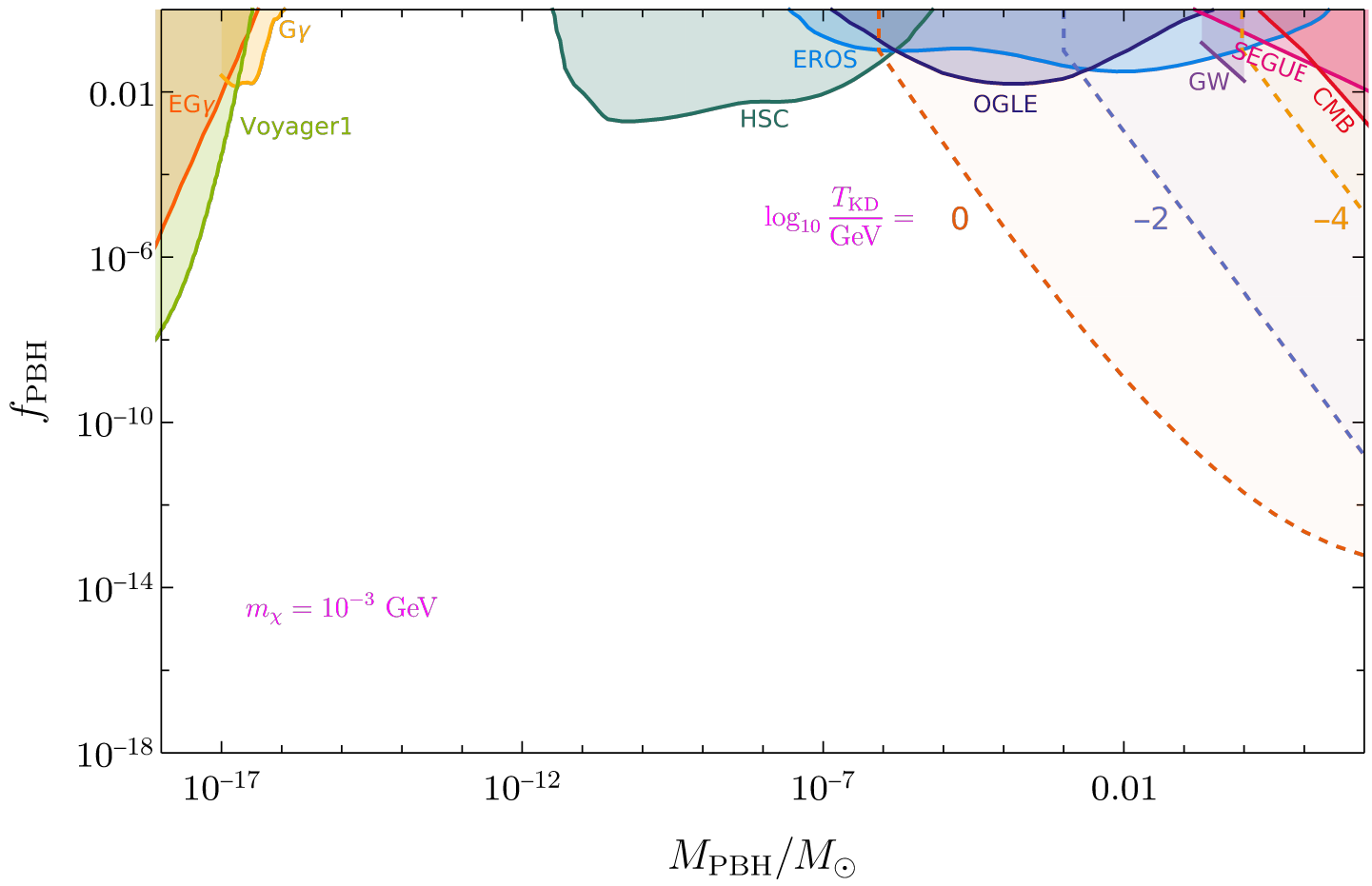}
    \includegraphics[width=0.49\textwidth]{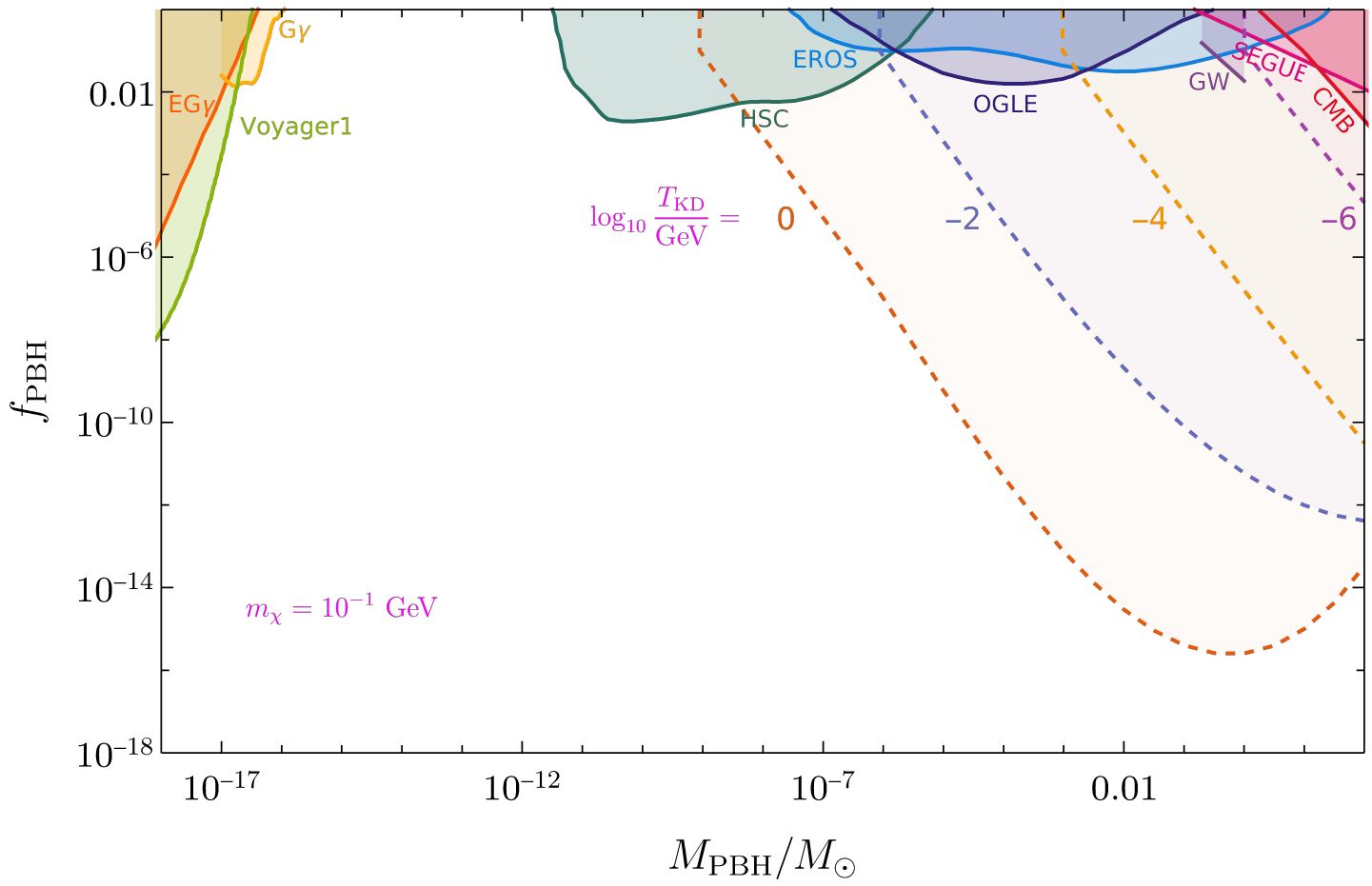}\\
    \includegraphics[width=0.49\textwidth]{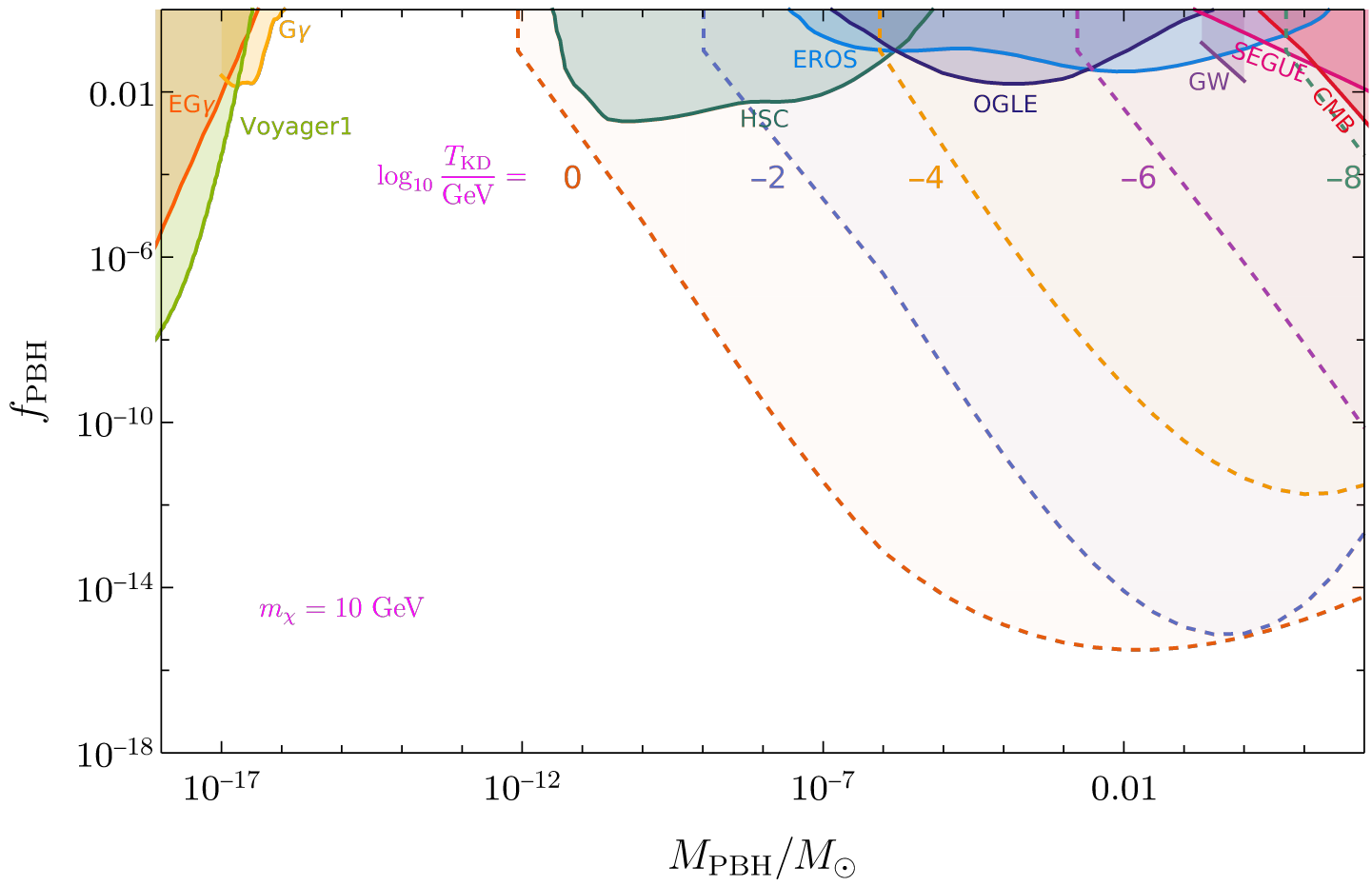}
    \includegraphics[width=0.49\textwidth]{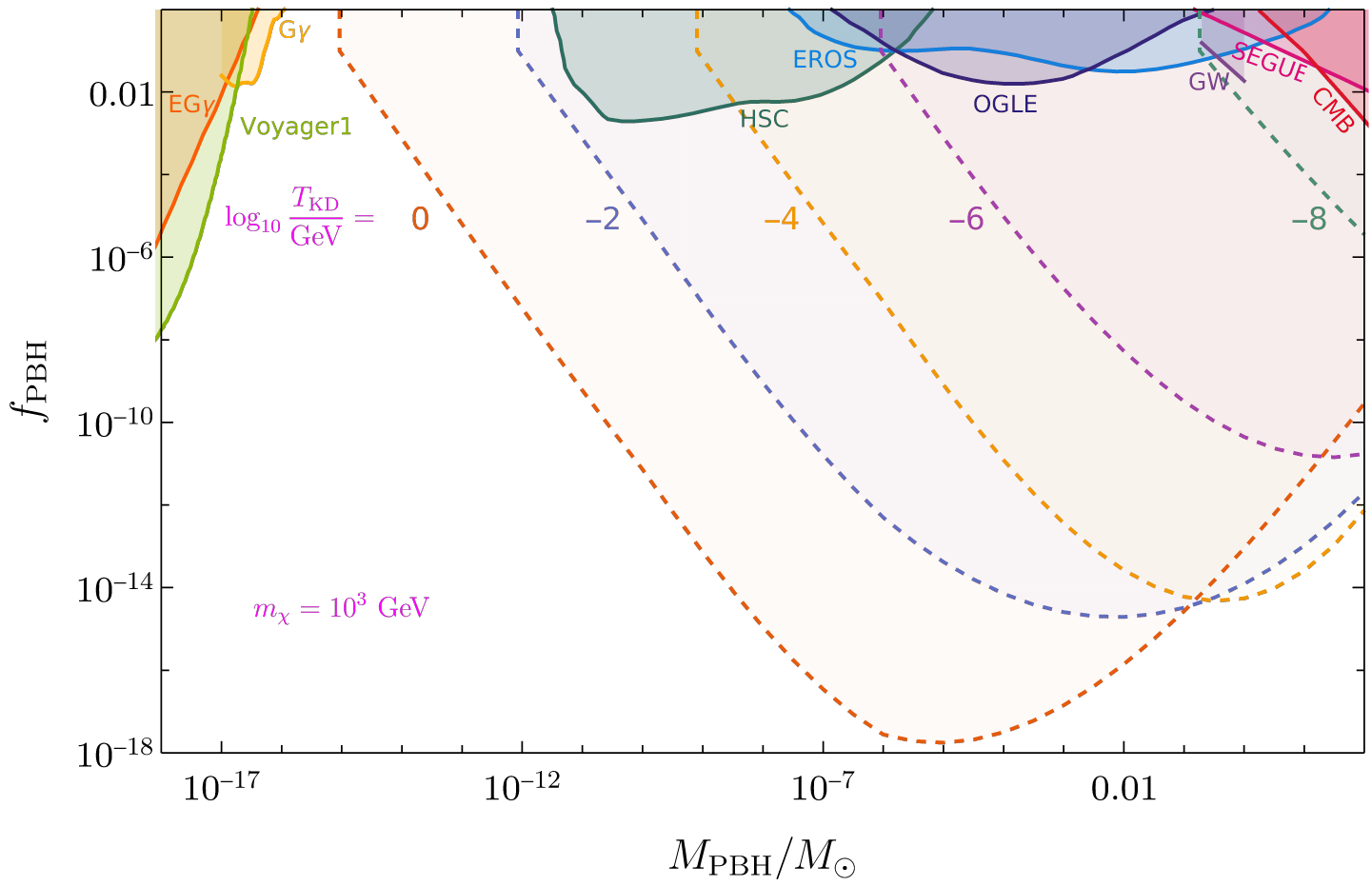}
    \caption[]{Constraints on the fraction of PBHs in DM $f_{\mathrm{PBH}}$ from the Galactic 511 keV gamma-ray line in mixed DM model with different mass $m_{\chi}$ and kinetic decoupling temperature $T_{\mathrm{KD}}$ of DM particle.
        We also show the constraints from the extragalactic gamma ray background (EG$\gamma$ \cite{Carr:2009jm}), galactic 511 keV line from Hawking radiation (G$\gamma$ \cite{DeRocco:2019fjq,Laha:2019ssq,Dasgupta:2019cae,Laha:2020ivk}), Voyager 1 measurements (Voyager1 \cite{Boudaud:2018hqb}), gravitational lensing events (HSC \cite{Niikura:2017zjd}, EROS \cite{Tisserand:2006zx}, OGLE \cite{Niikura:2019kqi}), gravitational waves (GW \cite{Authors:2019qbw}), dynamical effects (SEGUE \cite{Koushiappas:2017chw}), and cosmic microwave background (CMB~\cite{Poulin:2017bwe}). }
    \label{fig:con_fPBH_cosv}
\end{figure}

With the upper limit of $C_{\mathrm{D}}/C_{\mathrm{A}}$ obtained from 511 keV data in section~\ref{sec:data} and the density profile of halo around PBHs eq.~\eqref{eq:rho_chi_PBH}, the constraints on the abundance of PBHs in the total DM eq.~\eqref{eq:con_fPBH} can be written as
\begin{equation}\label{eq:con_fPBH_cosv}
    \begin{aligned}
        \frac{f_{\mathrm{PBH}}}{(1-f_{\mathrm{PBH}})^{2}} 
    &= \frac{2 M_{\mathrm{PBH}}}{\int_{0}^{r_{c}} 4\pi r^{2} dr
        \rho_{\mathrm{max}}^{2} + \int_{r_{c}}^{\infty} 4\pi r^{2} dr
    \rho_{\mathrm{b}}^{2}(r)} \frac{C_{\mathrm{D}}}{C_{\mathrm{A}}} \\
    &< \frac{3 M_{\mathrm{PBH}}}{2\pi \rho_{\mathrm{max}}^{2} r_{c}^{3} }
    \times 0.37\ \GeV/\cm^{3} .
    \end{aligned}
\end{equation}
Since the density profile outside of $r_{c}$ decreases fastly, we neglect the contributions from the parts out of $r_{c}$, which gives the last inequality of eq.~\eqref{eq:con_fPBH_cosv} and make the constraint more conservative.
And we apply the 95\% upper limit of $C_{\mathrm{D}}/C_{\mathrm{A}}$ for Einasto profile as it gives a better goodness of fit.
The constraints on the PBH abundance for different DM masses $m_{\chi}$ and DM kinetic decoupling temperatures $T_{\mathrm{KD}}$ are shown in figure~\ref{fig:con_fPBH_cosv}.

We cut the constraints at $f_{\mathrm{PBH}}=0.1$ because we have assumed $f_{\mathrm{PBH}} \ll 1$.
One can find that DM particle with larger mass and higher kinetic decoupling temperature will give more stringent constraints, which is due to that such particles can form denser halo around the PBHs and lead more emissions of 511 keV photons.
The constraints have $M_{\mathrm{PBH}}^{-2}$ slope in the relative small mass range which results from the fact that $r_{c}/r_{g}$ is about constant for PBHs with mass in the relatively small mass range as shown in figure~\ref{fig:rho_b}.

Noticing that these constraints are general and independent of particle DM models, which means that any particle DM model proposed to explain the 511 keV gamma-ray line observations will give such constraints on the abundance of PBHs.
For a given DM model, once one knows the mass and the kinetic decoupling temperature of DM particles, one can quickly get the constraints on the PBH abundance from figure~\ref{fig:con_fPBH_cosv}.

Among the scenarios proposed to explain the 511 keV gamma-ray line, there are two typical models, LDM and XDM.
For the LDM scenario, the low-energy galactic positrons are produced by direct annihilation of the LDM (about few MeV) particles into electron-positron pairs~\cite{Boehm:2003bt}.
For the XDM scenario, the DM particle has an excited state 1-2 MeV above the ground state, which may be collisionally excited and de-excited by positron-electron pair emission, which converts the kinetic energy of DM into positron-electron pairs~\cite{Finkbeiner:2007kk,Pospelov:2007xh}.
Although the positrons are not produced by directly annihilation in the XDM scenario, the positron production rate is also given by eq.~\eqref{eq:nea} in such a scenario, thus the analyses for the mixed DM consisting of annihilating particles and PBHs can also be applied to the XDM scenario.

\begin{figure}[htpb]
    \centering
    \includegraphics[width=0.7\textwidth]{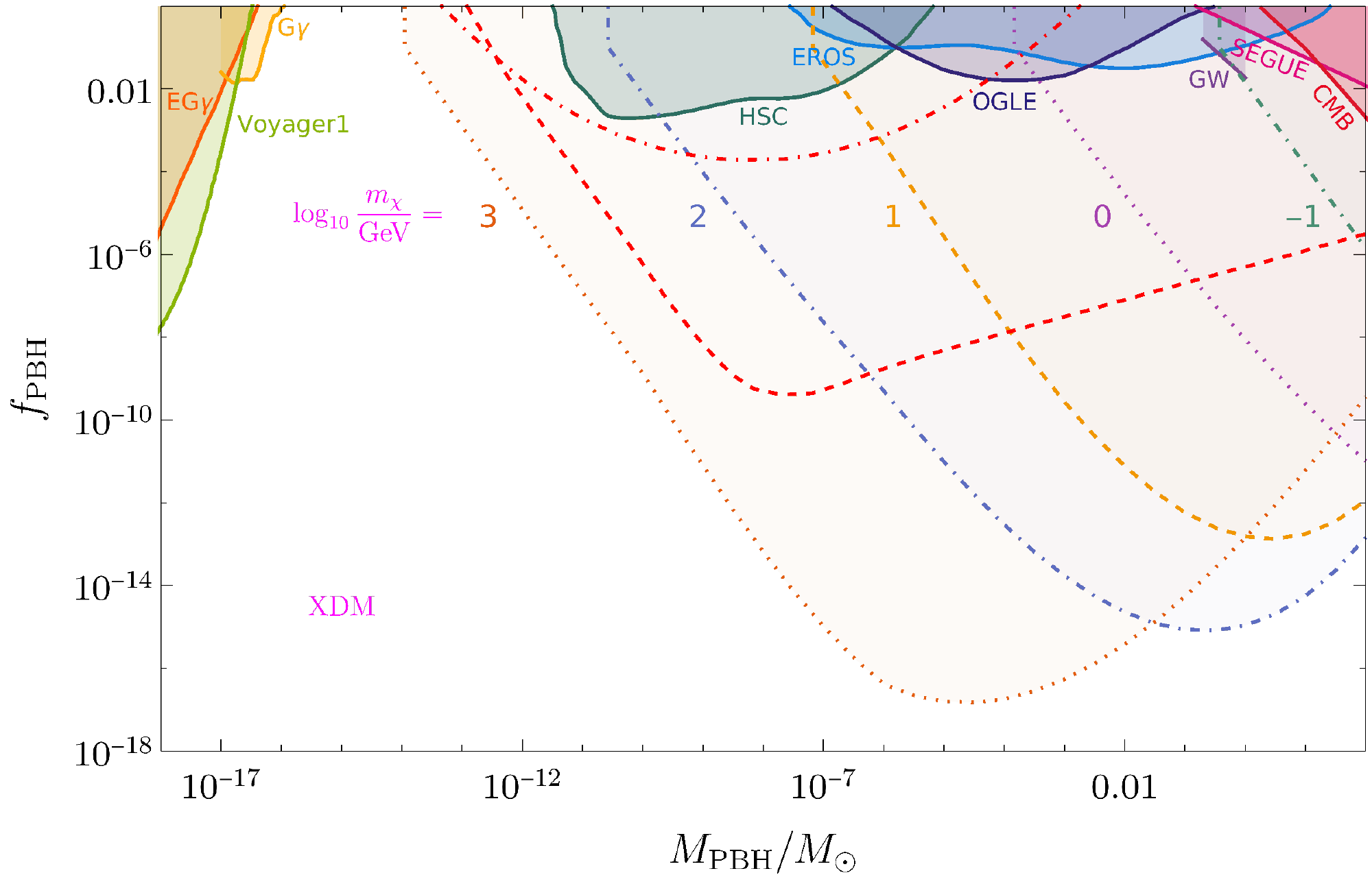}
    \caption[]{Constraints on the fraction of PBHs in DM $f_{\mathrm{PBH}}$ from the Galactic 511 keV line in the mixed DM model consisting of XDM and PBHs with different mass of DM particles.
        The red dashed line and red dot-dashed line indicate the constraints on the PBH abundance in the mixed DM model with $m_{\chi}=100~\GeV$ and $m_{\chi}=1~\TeV$ obtained from the extragalactic gamma-ray background, respectively~\cite{Boucenna:2017ghj}.
        The other constraints are same as shown in figure~\ref{fig:con_fPBH_cosv}. }
    \label{fig:con_fPBH_cosv_XDM}
\end{figure}

As a demonstration, we consider an XDM scenario and apply the relation between the kinetic decoupling temperature and mass for DM particles given by~\cite{Bringmann:2006mu,Bringmann:2009vf,Visinelli:2015eka,Boucenna:2017ghj}:
$
T_{\mathrm{KD}} = {m_{\chi}} \left( {\alpha m_{\chi}}/{M_{\mathrm{Pl}}} \right)^{1/4}{\Gamma(3/4)}^{-1}
$,
where $\Gamma(\cdot)$ is the gamma function, $\alpha=\sqrt{16\pi^{3} g_{*}(T_{\mathrm{KD}})/45}$ where $g_{*}(T)$ is the effective number of relativistic degrees of freedom.
Therefore the constraints on the PBH abundance in a mixed DM model with XDM and PBHs are dependent on the mass of DM particles.
We show the constraints with different mass of DM particles in figure~\ref{fig:con_fPBH_cosv_XDM}.
One can find that the constraints from 511 keV gamma-ray line observations can be down to $O(10^{-17})$ for XDM with $m_{\chi}=1~\TeV$, which are much more stringent than the constraints obtained from the extragalactic gamma-ray background (down to $O(10^{-9})$)~\cite{Boucenna:2017ghj,Adamek:2019gns}, and one should remember that these constraints are still conservative since we neglected the contributions of halo out of $r_{c}$ and did not consider additional decaying particle components as mentioned before.
On the other hand, such stringent constraints on PBHs means that the observation about PBHs can give a stringent constraint on DM explanations of 511 keV gamma-ray line observations and can help us to understand the origin of galactic 511 keV signal.

Considering the velocity dependence of the thermally averaged cross-section $\langle \sigma v \rangle$, we estimate the particle velocity distribution function of the Milky Way DM halo and the spike halo around PBHs, by using Eddington's formula~\cite{Lu:2017jrh,2008gady.book.....B}.
The results show the average particle velocity of the spike halo around PBHs is larger than that of the Milky Way DM halo for particles with $m_{\chi} \gtrsim 100 \GeV$, which could lead to more stringent constraints, and more details should be studied in future.

In this work, we get constraints from the data of INTEGRAL/SPI.
Assuming that the positrons annihilate close to their birth positions, then eq.~\eqref{eq:D_PBH} implies that there may be point sources of 511 keV emission.
There is a study searching for the 511 keV gamma-ray line from galactic compact objects with the IBIS gamma ray telescope~\cite{DeCesare:2011gc}, which can also be used to constrain this mixed DM model.
Due to the sensitivity of IBIS, current constraints from compact objects is looser than the one from INTEGRAL/SPI, but it has chance to give more stringent constraints in the future with the improvement of sensitivity.

\section{Conclusions}

The 511 keV gamma-ray line has been observed since 1970's but its origin is still not yet clearly known.
Apart from the astrophysical explanations, the possibilities that this signal comes from DM are also investigated extensively.
Among these DM explanations, it was found that DM decaying scenarios are disfavored by data, but this statement was obtained from data analyses which only consider single component DM models.
Since there is no evidence that DM is composed of only one component, we analyze the data of 511 keV gamma-ray line measured by INTEGRAL/SPI for the DM model consisting of both annihilating and decaying components.
Our work confirms the statement in earlier studies, and what more important is that we obtain the upper limit of the decaying components.

Except for particle DM, PBH is also an extensively studied candidate of DM.
With the existence of PBHs, the DM particles may be gravitationally bound to the PBHs and form halo around PBHs with density spikes.
These density spikes can enhance the production rate of positrons from DM particles, thus they can be  constrained by the observations of 511 keV line.

We consider a mixed model consisting of annihilating DM particles and PBHs, calculate the density profile of halo around PBHs, and get the constraints on the PBH abundance.
These constraints are general and independent of particle DM models, which means that any particle DM model proposed to explain the 511 keV gamma-ray line observations will give such constraints on the abundance of PBHs.
For a given DM model, once one knows the mass and the kinetic decoupling temperature of DM particles, one can quickly get the constraint on the PBH abundance from figure~\ref{fig:con_fPBH_cosv}.
For the mixed DM model consisting of XDM and PBHs, the constraints on the PBH abundance for DM particles with mass around $1~\TeV$ can be down to $O(10^{-17})$, which is much more stringent than that obtained from the extragalactic gamma-ray background.
These constraints are still conservative, and we expect the more stringent constraints can be obtained with more detailed studies and the improving sensitivity of experiments.

\section*{Acknowledgments}

We thank Shao-Jiang Wang and Yong Zhou for useful discussions.
RGC and XYY are supported in part by the National Natural Science Foundation of China Grants No. 11947302, No. 11991052, No. 11690022, No. 11821505 and No. 11851302, and by the Strategic Priority Research Program of CAS Grant No. XDB23030100, and by the Key Research Program of Frontier Sciences of CAS.
YCD and YFZ are partly supported by the National Key R\&D Program of China No. 2017YFA0402204 and by the National Natural Science Foundation of China (NSFC) No. 11825506, No. 11821505, No. U1738209, No. 11851303 and No. 11947302.

\bibliographystyle{JHEP}
\bibliography{004_ref}

\end{document}